\documentclass{article}
\usepackage{arxiv}
\usepackage[utf8]{inputenc} 
\usepackage[T1]{fontenc}    
\usepackage{hyperref}       
\usepackage{url}            
\usepackage{booktabs}       
\usepackage{amsfonts}       
\usepackage{nicefrac}       
\usepackage{microtype}      
\usepackage{lipsum}		
\usepackage{graphicx}
\usepackage[numbers]{natbib}
\usepackage{doi}

\title{NWP-based deep learning for tropical cyclone intensity prediction}

\author{\href{https://orcid.org/0000-0001-8947-8534}{\includegraphics[scale=0.07]{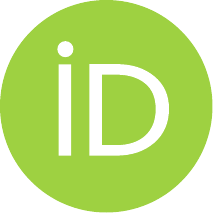}\hspace{1mm}Chanh Kieu}\thanks{Corresponding Author: ckieu@iu.edu} \\
	Department of Earth and Atmospheric Sciences\\
	Indiana University Bloomington, 47405, Indiana\\
	\And
        Khanh Luong \\
	Department of Earth and Atmospheric Sciences\\
	Indiana University Bloomington, 47405, Indiana\\
	\AND
        Tri Nguyen \\
	Luddy School of Informatics, Computing, and Engineering\\
	Indiana University Bloomington, 47405, Indiana\\    
}

\renewcommand{\shorttitle}{\textit{arXiv} Deep learning tropical cyclone intensity}

\hypersetup{
pdftitle={A template for the arxiv style},
pdfsubject={q-bio.NC, q-bio.QM},
pdfauthor={David S.~Hippocampus, Elias D.~Striatum},
pdfkeywords={First keyword, Second keyword, More},
}

\begin{document}
\maketitle

\begin{abstract}
Global artificial intelligence (AI) models are rapidly advancing and beginning to outperform traditional numerical weather prediction (NWP) models across metrics, yet predicting regional extreme weather such as tropical cyclone (TC) intensity presents unique spatial and temporal challenges that global AI models cannot capture. This study presents a new approach to train deep learning (DL) models specifically for regional extreme weather prediction. By leveraging physics-based NWP models to generate high-resolution data, we demonstrate that DL models can better predict or downscale TC intensity and structure when fine-scale processes are properly accounted for. Furthermore, by training DL models on different resolution outputs from physics-based simulations, we highlight the critical role of fine-scale processes in larger storm-scale dynamics, an aspect that current climate datasets used to train most global DL models cannot fully represent. These findings underscore the challenges in predicting or downscaling extreme weather with data-driven models, thus proposing the new role of NWP models as data generators for training DL models in the future AI model development for weather applications. 
\end{abstract}


%
%
\section{Introduction}\label{sec:introduction}
Tropical cyclone (TC) intensity, defined as the maximum 1-minute or 10-minute average of wind speed at the 10-m level (VMAX), is one of the key metrics measuring the strength of TCs \cite{BrownFranklin2002, Vukicevic_etal2014}. Obtaining this VMAX value in real TCs is however a challenging problem, as we do not often know exactly where the maximum wind location is due to the strong fluctuation of wind field from place to place. In the presence of satellite images, remote observation, and flight data, VMAX can be determined with a current accuracy of 2.5-4 m s$^{-1}$ (5-7 knots) with the current best retrieval algorithm and measurement \cite{Chen_etal2019, Wimmers_etal2019, TornSnyder2012, BrownFranklin2002, tian_etal2023}.  

For climate projections or real-time prediction of TC intensity, how to accurately derive TC intensity from a model gridded output is practical yet still elusive in current TC research. This is largely because TC intensity varies strongly with the model resolution, temporal frequency output, and approximations in computing the 10-m wind speed in traditional numerical weather prediction (NWP) models \cite{Gopal_etal2011, Vukicevic_etal2014, franklin_brown2008, Velden_etal2006, Vecchi_etal2013, Vu_etal2020}. The current approach for handling such an uncertainty in TC intensity is to use regional models at a higher resolution to resolve better TC intensity, which is applied to both operational TC forecasts and future projections in climate research \cite{Knutson_etal2007, Jourdain_etal2011, Redmond_etal2015, Kim_etal2015, Vu_etal2024}. Regardless of how high-resolution or frequent the output from an NWP model one may have, the current approaches all suffer from an issue of searching for VMAX on a gridded model output, which always underestimates the actual VMAX at any instance of time. That is, $\max_{\mathbf{x} \in G} |V(\mathbf{x})| \le \max_{\mathbf{x} \in R} |V(\mathbf{x})|$, where $V(\mathbf{x}), G, R$ are wind speed as a function of spatial coordinate $\mathbf{x}$, the model grid space, and the real atmospheric space, respectively.  

While using higher-resolution NWP models remains the best practice for addressing resolution-related challenges in regional models, improving TC intensity forecasts through this approach is often constrained by computational resources and slow progress in understanding convective-scale TC processes. Additionally, factors such as model errors, domain configurations, and the representation of physical parameterizations further hinder advancements in reducing TC intensity forecast errors and projection uncertainties, especially when compared to improvements in TC track and frequency forecasts \cite{Tallapragada_etal2015, du_etal2013, Kieu_etal2021}. As a result, achieving reliable TC intensity forecasts and estimations remains a significant challenge.

Recent advance in artificial intelligence (AI) opens a dramatic shift in weather and climate research, which is based on data-driven instead of physical-driven NWP models. At the short-medium range, many recent global AI models could demonstrate their promising performance in weather prediction, outperforming the best global NWP models such as the European Centre for Medium-Range Weather Forecasts (ECMWF) across metrics and criteria. Since the first introduction of the FCN model in 2022 \cite{Pathak_etal2022}, a range of global AI models have been developed, with the most recent GenCast model clearly showing better performance than the best NWP models ECMWF \cite{Bi_etal2022, Lam_etal2023, Weyn_etal2019, Rasp_etal2020, Wang_etal2021predrnn, Chen_etal2023fengwu, Price_etal2025}. Such a rapid advancement of AI techniques for weather prediction applications suggests that these global models may soon overtake the traditional NWP models in operational global weather prediction, at least in terms of the global metrics that they work well. 

While global AI models can outperform NWP models in a range of weather metrics, at the regional scales of extreme weather events, the picture is different. This is because most extreme phases of weather such as heavy rainfall, thunderstorm activities, mesoscale convective systems or TC intensity vary quickly in time and widely in space at small scales that global AI models cannot capture. One clear example is VMAX, which is measured by the point-like maximum wind near the surface. For this metric, the current intensity forecast skill by all global data-driven models is in fact not yet comparable to any NWP models in terms of mean absolute errors, even when compared to the global NWP model at the 0.25-degree resolution \cite{Kieu2024b}. 

Such a worse performance of global AI models in predicting or downscaling TC intensity is due to several reasons. First, all current climate datasets used for training global AI models do not capture the actual TC structure corresponding to an observed TC intensity. Taken, e.g., the ERA5 hourly data at 0.25-degree resolution \cite{ERA5, Rasp_etal2020}, which is considered to be the best data for AI model training. While ERA5 contains all meteorological variables at many different levels and covers a long past period, this 0.25-degree resolution is simply too coarse to resolve any TC inner-core structure, especially at a very high-intensity state for which the eye size is around 30-50 km in radius. As such, trying to match ERA5 gridded data with the best-observed intensity from the best track is just beyond what AI models can handle. One may argue that there may be other datasets with finer resolutions such as satellite data or radar observations that can go along with best track TC intensity. However, these datasets do not cover all levels, required variables, or as many TCs as expected for training a typical AI model. To the best of our knowledge, there is currently no observed TC datasets at sufficiently high resolution that can match a proper TC structure with the observed TC intensity that is reported in the best track database.  

Second, all AI global models are built on a loss function that emphasizes the atmospheric global state instead of the local state for extreme events during training. Such a global focus is critical for global AI models, because these models aim to capture the large-scale evolution of the atmosphere, not the spare extreme weather events at local points that occur infrequently. In the case of TC intensity, this is an important issue as the global loss function is simply not designed for such occasionally very high-intensity values at a few isolated points. One may want to introduce an additional loss function term to account for TC intensity or other extreme events to improve intensity forecast skill. However, the rarity of these extreme values means that any such additional loss function will be dominated by the global loss function eventually during the training. 

Note that these above issues with data resolution and global designs are not limited to TC intensity but applied to all other extreme weather features for which we do not have good observational data to capture their detailed structures. Events such as heavy rainfall, mesoscale convective systems, severe thunderstorms, or tornadic events all possess a local and brief phase of maximum strength in time that no current reanalysis datasets can include. Given these issues and the available climate data for AI model training, an apparent question is whether AI models can be developed for these regional extreme weather purposes beyond the traditional NWP or statistics/dynamical downscaling approach still largely remains. 

Specific in this study, we highlight the need for regional AI models to incorporate the ability to track TCs and account for lateral boundary conditions, similar to regional NWP models \cite{Kieu2024b}. Unlike traditional object detection algorithms in image processing, extreme weather systems such as TCs often evolve rapidly under the influence of steering flows and environmental conditions entering through the model’s lateral boundaries. This boundary information introduces additional complexities in developing regional AI models, beyond challenges related to data resolution and loss function issues, as discussed earlier

Given the importance yet inherent challenges of regional AI models in predicting extreme weather events, there is a clear need for an approach that can enhance the performance of AI models in this domain. In this study, we present a general framework that leverages physics-based models to generate high-resolution data for various types of extreme weather events, which can be used to support AI model development. Our primary focus is on TC intensity, a complex phenomenon characterized by multi-scale processes and rapid intensity fluctuations over time. The core motivation for our approach lies in the absence of comprehensive observational datasets capable of capturing TC structures across their phases of development. To address this gap, we utilize high-resolution NWP models to generate training data with detailed outputs and high fidelity, which are essential for developing robust AI models. To some extent, our approach of employing NWP models to produce synthetic training data shares similarities with generative adversarial networks. However, the context and methodology of our application are more specifically tailored to address local extreme-weather events in this study. 
%
%
\section{Deep-learning model architecture}\label{sec:model}
With the focus on TC intensity and structure retrieving from coarse-resolution gridded data, it is important to have a good deep-learning (DL) architecture that could extract TC information effectively from gridded data. In this study, we will use the convolutional neural networks (CNN) architecture to predict TC intensity and structure. Previous applications of CNN models for TC research show that CNN can extract information from input data if a proper design and hyperparameters are set \cite{Wimmers_etal2019, Chen_etal2019, tian_etal2023, NguyenKieu2024}. Because of its capability, the CNN architecture is chosen to retrieve TC information from gridded data in this study.       

For our specific application, we started from a base CNN architecture for TC intensity (hereinafter referred to as TCNN) that consists of five convolutional layers, with kernel sizes of 32, 64, 128, 256, and 512, respectively. Each convolutional layer in the network, except for the last, employs Rectified Linear Unit (ReLU) activation, a dropout rate of 0.1, and \texttt{same} padding to ensure non-linearity and maintain spatial dimensions throughout the processing stages. Note that the last convolutional layer utilizes \texttt{valid} padding, aimed at reducing the output size to focus on the most relevant features. Strides are fixed at 1$\times$1 for convolutional layers and 2$\times$2 for max-pooling layers. A total of three max-pooling layers are used after the first three convolutional layers. The output layer consists of a single unit, also using \texttt{ReLU} activation, configured for regression-based predictive tasks. The model employs the Huber loss function and Adam optimization. 

Among all of these hyperparameters, we notice that the input data domain size turns out to be the first important parameter in our problem, which needs to be tuned for each model resolution. In our configuration, the default setting with input domain sizes of 201$\times$201 and 23$\times$23 are applied for the 2-km and 18-km resolution input data, respectively. These domain sizes ensure that the input data covers the same physical domain for both resolutions. With five channels, these input data domains are then resized to the same number of pixels in the $(row,column)$ dimensions such that both domains for the 18-km and 2-km resolution can have the same data shape of 64$\times$64$\times$5 before training.  

The second important hyperparameter is the number of CNN layers, which should not be too large (5 in our design) to avoid the vanishing gradient problem. One could address the vanishing gradient issue by using a deeper network and applying some skipping mechanisms such as those used in ResNet models to overcome this issue \citep{He_etal2016}. However, for the current problem of extracting TC features for intensity retrieval at the highest resolution of 2 km, our experiments show that the above design suffices for extracting TC intensity without all the complications of training a deeper convolutional network. An optimal number of CNN layers likely depends on each model resolution, which requires re-training a DL model properly.       

The entire workflow and executions were carried out using the TensorFlow framework, which supports a variety of model architectures, utilities, and designs. The TCNN model was trained using both the root mean square error (RMSE) and the mean absolute error (MAE) used for the accuracy metric. Due to the high-resolution training data, we restrict our batch size to 256, with 1000 epochs for all training. In addition, we used the \texttt{Adam} optimizer with a learning rate of $10^{-3}$ and a decay rate of 0.0001 that helps speed up the training process. A Keras callback utility was used for the early stopping, which saves the best model during the training based on the validation data. The whole training of the TCNN model on an A100 GPU took $\approx$ 30-180 minutes, depending on the resolution of the input data.  
%
%
\section{Data-generation description}\label{sec:description}
\subsection{WRF configuration}\label{sec:wrf}
To generate very high-resolution data that can be used to train the TCNN model, we use the Weather Research and Forecasting (WRF, Version 3.9 \cite{WRF2019}) model in this study. With the goal of maximally isolating the impacts of data resolution on TC intensity retrieval, we employed the idealized configuration of the WRF model such that the model vortex has the same environmental conditions in all experiments. Our idealized experiment was designed with triple nested domains at resolutions of 18, 6, and 2 km whose corresponding grid size dimensions are 301$\times$301, 20$\times$201, and 201$\times$201, respectively. For the vertical dimension, we configured the model with 41 sigma levels and a model top of 25 km. 

With those settings, all WRF idealized simulations in this study were initialized on an $f$-plane at 20N latitude and constant sea surface temperature of 302K, using the environmental sounding from the Dunion’s sounding. Similar to \cite{Kieu_etal2014}, we implemented an initial vortex in the WRF model using the wind profile in the radius-height coordinate derived from the previous study \cite{KieuZhang2009} as follows:
\begin{equation}
V(r,\sigma) = V_m \frac{r}{r_m} sin (\frac{\pi}{2} \sigma)^{1-\delta}cos(\frac{\pi}{2}\sigma)^{\delta}e^{\frac{1}{b}[1-(\frac{r}{r_m})^b]}    
\end{equation}	
where $V_m$ is VMAX, $r_m$ is the radius of the maximum wind (RMW), $b$ is the nondimensional parameter determining the horizontal shape of the wind profile set to be 0.7 in our experiment, $\delta \in [0,1]$ is a nondimensional parameter controlling the height of VMAX, and $(r,\sigma)$ is the radial and vertical coordinates. Note that $\delta$ is needed because TC initial structure during their early stage is often around mid-troposphere. As such, adjusting $\delta$ can give us some control over different initial structures of the model vortex as discussed in \cite{Kieu_etal2014}.
%
%
\begin{figure}[ht]
\centering
\includegraphics[width=13.5cm]{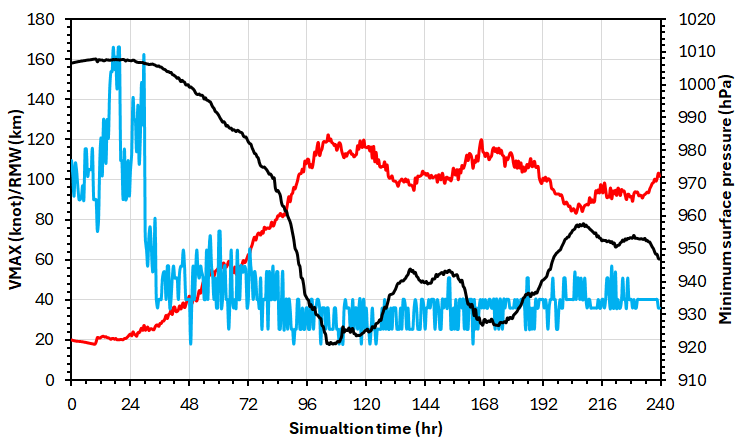}
\caption{A time series of VMAX (knots), the minimum central pressure (PMIN, hPa), and the radius of maximum wind (RMW, km) from a 10-day simulation as obtained from the WRF model simulation. Note that the unit for VMAX is in knot (1 knot $\approx$ 0.5 m s$^{-1}$) to facilitate all time series on the same panel.}
\label{fig:time_series}
\end{figure}

With those model settings, all WRF idealized simulations in this study were integrated for 10 days, with an output frequency of every 10 minutes. Figure \ref{fig:time_series} confirms that the model quickly settles down to a maximum intensity state after 4 days into integration. To prevent the training from skewing toward strong TC intensity states, we limited all of our simulations to 10 days. 

To produce different TC structures and intensity at the same resolution for DL training, we employed a range of model physics for the WRF simulations in generating our TC dataset. Specifically. we used 4 microphysical schemes including Kessler, Lin et al, WSM 5-class, and WSM-6-class, two long-wave radiative schemes: RRTM and RRTMG, two boundary layer schemes: YSU and Mellor-Yamada-Janjic, and two cumulus parameterization schemes for domains 1 and 2 including Kain-Fritsch (new Eta) and modified Tiedtke. All simulations had periodic boundary conditions in the west-east and north-south directions for the outermost domain. 

\begin{figure}[ht]
\centering
\includegraphics[width=13.5cm]{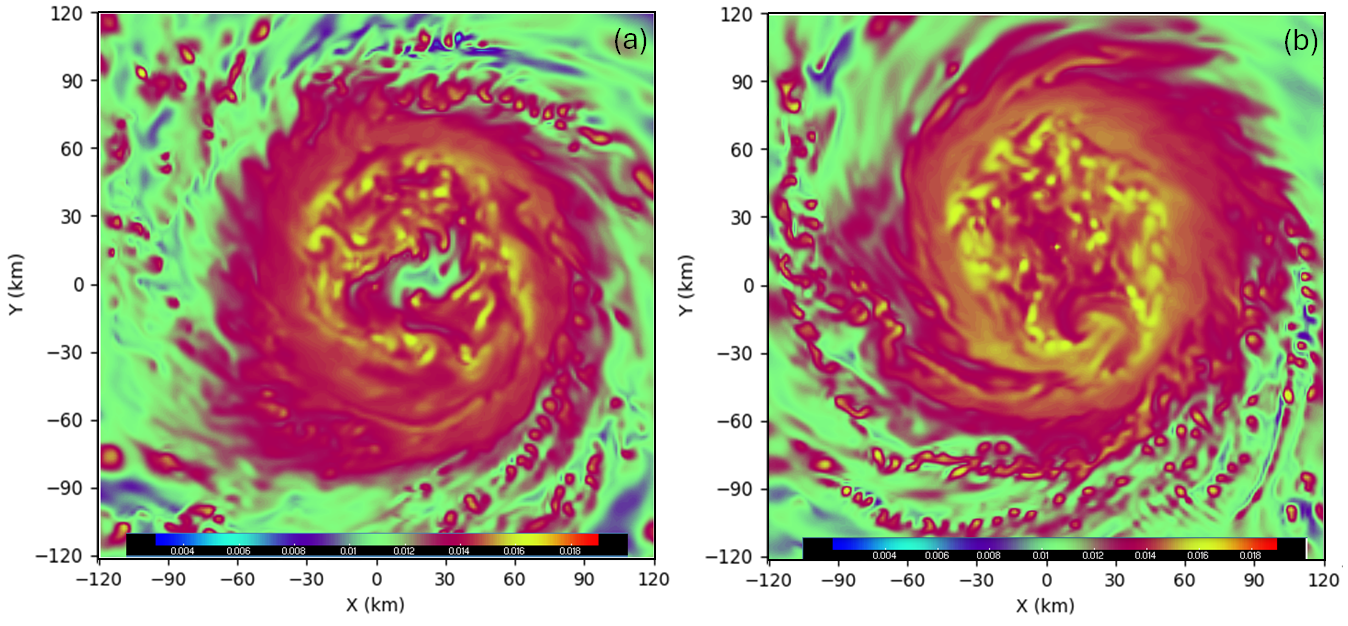}
\caption{A snapshot of the horizontal cross section of 850-hPa specific humidity (shaded, unit kg kg$^{-1}$) at two different moments $t=118$ hr and $t=162$ hr for which the WRF model possesses the same VMAX as shown in Fig. \ref{fig:time_series}.}
\label{fig:structure_same_Vmax}
\end{figure}

We note that the use of diverse physical schemes to generate the high-resolution dataset is a critical aspect of our approach. Each physical scheme offers unique advantages and disadvantages when simulating TCs, and to date there is no universally optimal scheme capable of accurately representing all TCs across all ocean basins. This variability also reflects the inherent nature of TC intensity, with no two TCs sharing identical structures at the same VMAX (Fig. \ref{fig:structure_same_Vmax}). In other words, a given intensity can correspond to multiple TC structure during real TC development. Consequently, using multiple physical schemes in our WRF simulations should be viewed as a data augmentation strategy, generating a variety of TC structures for training DL models within an idealized framework.
%
%

\subsection{Experimental design}\label{sec3.2}
To focus on our main aim of examining the impacts of high-resolution data for training DL models, we designed two sets of WRF simulations in this study. In the first set of simulations, we conducted a range of triple-nest domain simulations with the same initial condition and domain configurations, but using different physical parameterizations. These simulations (hereafter referred to as H02) were run for 10 days, with an output frequency of every 10 minutes to capture all phases of TC development. Note that one can always run the WRF model as long as one wishes. However, the model always quickly settles down to a maximum intensity limit after just 4 days into integration, and running a longer simulation does not help the DL training in terms of data representation for different stages of TC intensity (Fig. \ref{fig:time_series}). The outcome from this H02 experiment is a dataset of 14,400 data points (i.e., ten 10-day simulations at the 10-minute frequency). 

In the second set of simulations, identical setups and the number of simulations as for the H02 experiment were conducted, except we used only 1 domain at 18-km resolution. This set of lower-resolution simulations (hereafter L18) is needed to demonstrate the effects of fine-scale processes on TC development and DL model performance. For example, using TC structure data from the L18 dataset and TC intensity from the 2-km resolution in the H02 dataset for training DL models, one can demonstrate what data-driven models can best achieve from the low-resolution input data. This is very similar to the practical situation where the best track TC intensity is derived from satellite observation, while the TC gridded data is provided by a coarse-resolution model output. 

Likewise, by comparing our TCNN model performance using the 18-km resolution data from the L18 and H02 datasets, one can further quantify the importance of fine-scale processes that are manifested at a coarser resolution grid. We note again that for the H02 experiment, all three nested domains have the feedback between a nest and its parent domain turned on. So, the 18-km output from H02 simulations contains some information from TC fine-scale physics/dynamics that the 18-km output from the L18 simulations cannot contain. Therefore, using H02 and L18 datasets generated by the WRF model can provide a complementary picture of the capability of DL models in extracting TC intensity from gridded data as expected. 

By default, the output from the WRF simulations contains many different key variables and levels. For our purpose of training the TCNN model to predict TC intensity and structure, we however extracted a subset of variables and levels from the WRF output and treated them as input channels for our TCNN model. Specifically, we selected relative humidity, two wind components ($u,v$), and temperature at 950, 850 and 250 hPa. In addition to these three-dimensional variables, the surface pressure and 500-hPa geopotential height were also used, creating a total of 14 input channels. Note that all of these variables have different dimensions and vary with height. Therefore, we need to normalize all of them before training, using max/min normalization.  

With the H02 and L18 datasets, we set aside one randomly selected simulation from each dataset for testing and split the remaining data into training and validation sets in a 95:5 ratio. Rather than following the traditional approach of combining all data and then splitting it, we chose this approach to prevent the overfitting of the TCNN model to a specific simulation. This concern is specifically relevant to TC intensity, which exhibits temporal dependencies. Training a DL model using both training and test data from the same simulation can introduce an inherent memory effect, making the model less generalizable and overly confident in cases beyond the training data.

For training and verification of the TCNN model, the label values including VMAX, PMIN, and RMW are directly computed from the model output. Notably, in our design, the training data (i.e., $X$ data) can be taken from one resolution, while the label data (i.e., $y$ data) can come from a different resolution. This approach allows us to not only evaluate the TCNN model’s performance but also quantify the impact of fine-scale processes on coarser resolutions. For instance, training on 18-km data while using label data from the 2-km resolution replicates the real-world scenario where actual TC intensity is obtained from the best-track data, while TC structural information is available only at a coarser resolution as mentioned above. This design is reflected in Table \ref{tab:experiments}, where the first three characters of a DL experiment acronym denote the source of the training data, and the last three characters indicate the label data source. Table \ref{tab:experiments} summarizes all WRF simulation settings and related DL experiments in this study.

While our TCNN model is a regional DL-based model for downscaling TC intensity, we wish to note that this model does not require the incorporation of lateral boundary conditions, as discussed in \cite{Kieu2024b}. Unlike spatiotemporal prediction problems for storm-following domains, which critically depend on lateral boundary information to forecast the evolution of a system in the next frame, downscaling TC intensity from a given gridded dataset inherently includes all lateral boundary conditions at every frame from input data. Therefore, no separate treatment of lateral boundaries is necessary.  

\begin{table}[ht]
\centering
\caption{WRF simulations and related DL experiments using the TCNN model. Note that the acronyms for the DL experiments follow a convention of $XNNynn$, where $XNN$ denotes the gridded training data on a given resolution $NN$ km, $ynn$ is the scalar label data including VMAX, PMIN, and RMW extracted from a resolution $nn$ km, and $X,y \in \{H,L\}$ denote either high resolution (H02) or low resolution (L18) simulations.}
\label{tab:experiments}
\begin{tabular}{p{1.8cm} p{11cm}}
\hline
Experiments &  Remarks\\
\hline
L18 & WRF simulations using a single domain at a low 18-km resolution, along with a range of model physical parameterizations scheme for data augmentation. \\
H02 & Identical to the L18 experiments except for using a triple-nested domain at higher resolution of 18/6/2-km.  \\
H02h02 & TCNN training using 2-km resolution data from the H02 experiment, and the label data (VMAX, PMIN, RMW) from the 2-km resolution data in the H02 experiment\\
H18h18 & TCNN training using 18-km resolution data from the H02 experiment, and the label data (VMAX, PMIN, RMW) also from the 18-km resolution data in the H02 experiment\\
H18h02 & TCNN training using 18-km resolution data from the H02 experiment, and the label data (VMAX, PMIN, RMW) also from the 2-km resolution data in the H02 experiment\\
L18\textit{l}18 & TCNN training using 18-km resolution data from the L18 experiment, and the label data (VMAX, PMIN, RMW) also from the L18 experiment\\
L18h02 & TCNN training using 18-km resolution data from the L18 experiment but the label data (VMAX, PMIN, RMW) from the 2-km resolution data in the H02 experiment\\
\hline
\end{tabular}
\end{table}

%
%
\section{Results}\label{sec:results}
\subsection{Deep-learning TC retrieval}
%
%
\begin{figure}[ht]
\centering
\includegraphics[width=13.5cm]{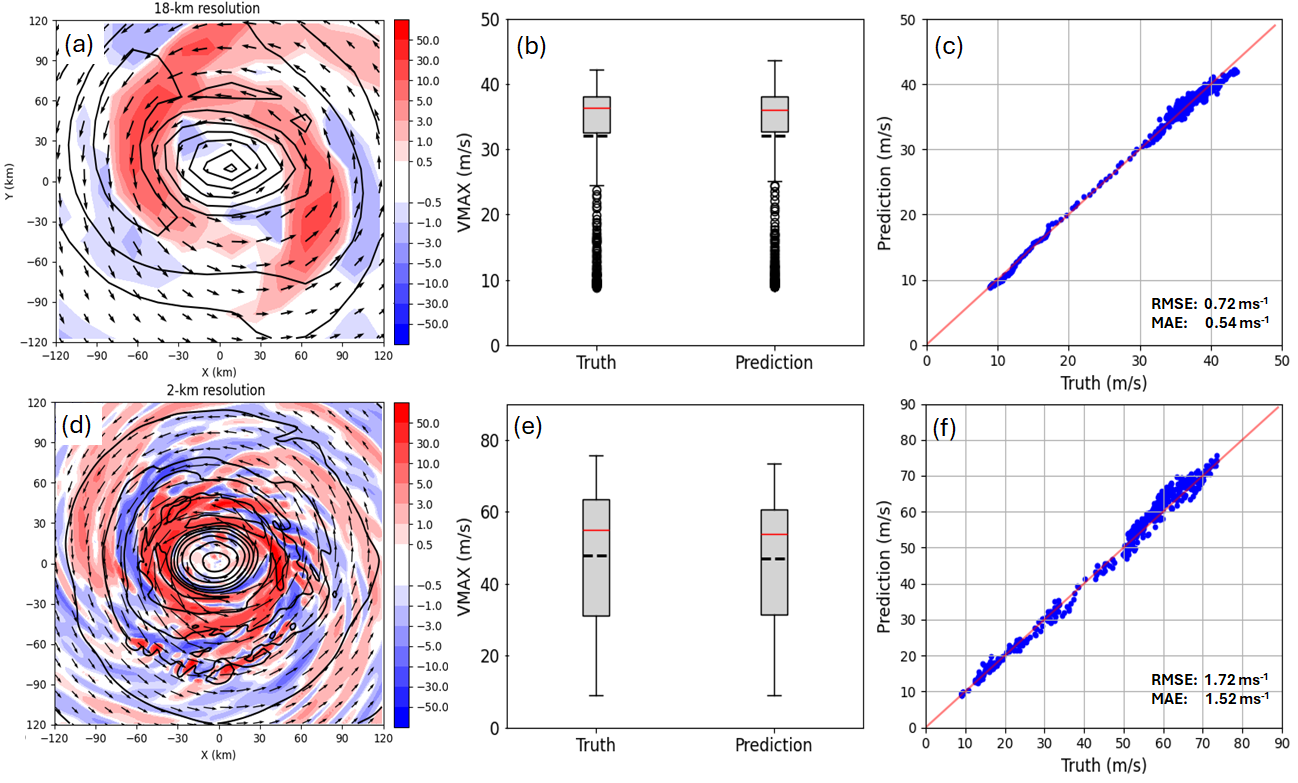}
\caption{(a) Horizontal cross section of vertical motion (shaded, unit 0.1 m s$^{-1}$) at the 850 hPa level valid at 8-day into integration and the corresponding tangential wind (contoured at an interval of 5 $m s^{-1}$) and flow field at the same level (vectors) obtained from the 2-km resolution data in a H02 simulation, b) a box plot of the VMAX (m s$^{-1}$ prediction from the TCNN model for the test set and the true values in the H02h02 experiment, and c) a scatter plot of VMAX prediction relative to the true values. (d)-(f) similar to (a)-(c) but for the lower-resolution experiment L18l18 that uses data from 18-km resolution simulation for training and prediction. Red lines in panels (b) and (e) denote median values, and red lines in panels (c) and (f) denote the perfection forecast.}
\label{fig:vmax_benchmark}
\end{figure}

To have a broad overview of the differences in input data generated from high- and low-resolution simulations used to train our DL models, Fig. \ref{fig:vmax_benchmark} presents a snapshot of 850 hPa wind channels from the 18-km domain in a 18-km resolution simulation and the 2-km domain in a 2-km resolution simulation. All simulations are conducted using the Weather Research and Forecasting (WRF) model, which is a common physics-based model for weather research, as outlined in the Methods section.

A notable distinction between the two resolution datasets is evident, with the 2-km dataset capturing significantly more detailed patterns in the inner-core region such as deep convective bands or high-wind streaks. While the precise location and magnitude of these fine-scale features are not generally predictable, their presence is itself crucial, as they are only visible in the 2-km resolution in H02 simulations. For DL model development, such features are important as they may help enhance the ability of DL models in predicting TC intensity. 

We emphasize that these fine-scale structures in the 2-km data as shown in Fig. \ref{fig:vmax_benchmark}a,c is not merely a visual difference between low and high-resolution images. These features actually represent the physical processes that can only develop at the convective scale when the WRF model is run at a sufficiently high resolution. Therefore, simply enhancing low-resolution data through traditional image processing techniques will not reveal or capture these fine-scale features, as they do not exist in lower-resolution simulations. 
How these fine-scale structures contribute to the performance of DL models in predicting or downscaling TC intensity is however unknown, which is what we wish to quantify in this study.       

With such detailed differences between low- and high-resolution data, we first present two DL experiments to benchmark our TCNN model for TC downscaling, which is based on convolutional neural networks. In the first experiment (H02h02), the 2-km resolution model output from the H02 dataset is used for training (i.e., $X$ data), while the target values (i.e, $y$ data) for TC intensity and structure are directly obtained from searching the maximum/minimum values on the 2-km gridded data in the H02 simulations. Here, we will follow the traditional measures that use three scalar metrics as proxies for TC intensity and structure, which include VMAX, and the minimum surface pressure (PMIN), and the radius of maximum wind (RMW). These metrics are treated as targets in our DL model training, which can be obtained from the WRF model output. 

The second experiment (L18l18) is similar to the H02h02 experiment, except that the 18-km resolution data is now used for both training and target values (see Table \ref{tab:experiments} for DL experiment designs and acronym convention). Both of the H02h02 and L18l18 experiments are needed to establish a benchmark for our TCNN model, because any effective DL model should be able to accurately recover TC intensity and structure from the input data using the target values on its own grid. Any difference between these experiments will underscore the net impact of model resolution on retrieving TC information by DL models.

Figure \ref{fig:vmax_benchmark} shows the VMAX prediction of the TCNN model from the H02h02 and L18l18 simulations, which are verified against the target values directly computed from a test set in 2-km and 18-km grids. One notices that the TCNN model can indeed retrieve well VMAX with the idealized data for both experiments with the root mean squared errors (RMSE) of 1.22 and 0.73 m s$^{-1}$ for the H02h02 and L18l18 experiments, respectively. As can be seen from both the box and scatter plots, the VMAX prediction by the TCNN model could capture all major statistics including mean, median, quartiles, and correlation as compared to the true values, even with the 18-km resolution training data.   

The above results are very noteworthy if one recalls that the typical RMSE for the currently best VMAX retrieval in operational practice is $\approx$3 - 4.5 m s$^-1$ ($~$6-9kts, \citep{Chen_etal2019, Wimmers_etal2019,tian_etal2023}). Here, our TCNN model shows a much better result with its RMSE from both the 2-km and 18-km data comparable to the natural noise level of VMAX fluctuation under idealized conditions (see, e.g., \cite{Kieu_etal2014}). Of course, any comparison of the TCNN model with real-time TC intensity retrieval is not fair, as our VMAX retrieval is carried out on a gridded dataset in idealized settings, while the real-time VMAX retrieval is based on satellite data and verified against flight or surface data. However, the great performance of the TCNN model in recovering TC intensity on its own model grid could at least confirm our model's capability under idealized conditions as designed.   

Of also interest is that retrieving VMAX in the lower resolution experiment L18l18 (RMSE of 0.73 m s$^{-1}$) is substantially more accurate than that in the higher resolution experiment H02h02 (RMSE of 1.72 m s$^{-1}$). At face value, this indicates that too detailed features in the higher resolution data from 2-km simulations tend to confuse the TCNN model due to stronger VMAX fluctuations at high resolution. Training any DL model with noisier data is generally more difficult, thus preventing the TCNN model to learn the right TC structure corresponding to a given VMAX value, even on its own grid. This explains for the higher RMSE in the H02h02 experiment as seen in Fig. \ref{fig:vmax_benchmark}.  

%
%
\begin{figure}[ht]
\centering
\includegraphics[width=12cm]{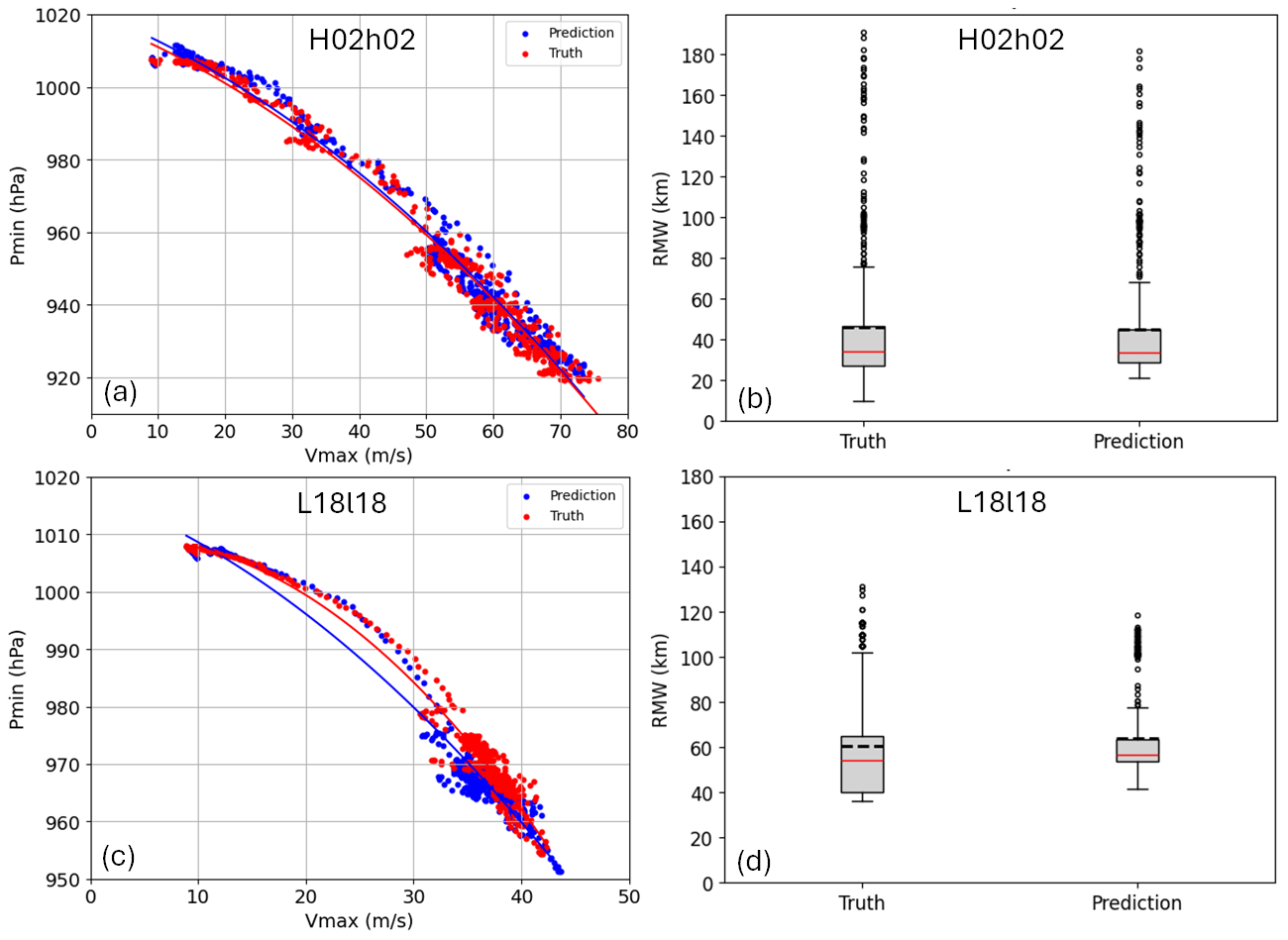}
\caption{(a) The pressure-wind relationship obtained from the H02h02 experiment (blue dots) and the actual relation obtained from the H02 simulations on the 2-km resolution grid (red dots), and b) a box plot of RMW (km) prediction as obtained from the H02h02 experiment for the test set versus the true distribution of RMW obtained from the H02 simulations. (c)-(d) similar to (a)-(b) but for the lower resolution experiment L18l18.}
\label{fig:pminRmw}
\end{figure}
In contrast, using PMIN as a proxy for TC intensity provides a different picture of the TCNN model performance between the H0202 and L18l18 experiments. This is best seen in the form of a pressure-wind relationship (PWR), which is a good measure of the model's dynamical constraint. Consistent with the VMAX retrieval, the TCNN model could predict well PMIN and related PWR (Fig. \ref{fig:pminRmw}a,c), with the RMSE of 3.47 hPa and 3.81 hPa for the H02h02 and L18l18 experiments, respectively. Notice that the RMSE for PMIN is now slight lower in the H02h02 experiment as compared to that from the L18l18, which is related to the fact that PMIN is generally more predictable. Unlike the strong fluctuations of VMAX from local wind streaks at very high resolution that can degrade the TCNN model's ability to learn, PMIN varies less rapidly with time. Thus, the TCNN model can retrieve PMIN with lower RMSE at higher resolution.  

In particular, the PWR obtained from the H02H02 experiment exhibits a better fit with the true PWR compared to that from the L18L18 experiment. While the weaker and smoother TC intensity values in the low-resolution grid allow the TCNN model to perform better for VMAX and PMIN individually, it fails to capture their relationship in cases with the highest VMAX and lowest PMIN in the L18L18 experiment. For instance, TCNN may predict a large VMAX but an insufficiently low PMIN, or vice versa. Consequently, the PWR in the L18L18 experiment shows a significant deviation from the true PWR. In contrast, in the H02H02 experiment, the TCNN model accurately captures the PWR across the entire range of TC intensity. This highlights how higher-resolution data help better constrain the model dynamics beyond individual intensity metrics.     

Along with its ability to constrain TC intensity and dynamics, the TCNN model also demonstrates skill in predicting RMW (Fig. \ref{fig:pminRmw}b,d). It is important to note that RMW differs significantly between the 2-km and 18-km resolution simulations, as higher resolution leads to finer vertical and horizontal structures during TC development, generally resulting in a smaller RMW (see Fig. \ref{fig:vmax_benchmark}a,d). At 2-km resolution, the RMW may reflect local wind gusts rather than the storm-scale TC structure, exhibiting fluctuations similar to those observed in VMAX. This is evident in Fig. \ref{fig:pminRmw}c,d, where the H02H02 experiment shows a lower mean and quartile in the RMW distribution, along with more outliers.

Corresponding to such a smaller and more variable RMW, the TCNN model exhibits a slightly higher RMSE for RMW in the H02H02 experiment (17.88 km) compared to the L18L18 experiment (13.77 km). This underscores the sensitivity of structural metrics like RMW to model resolution and suggests that very high-resolution data may introduce random fluctuations that hinder full optimization. These findings present an interesting insight: higher resolution data does not always yield better performance for certain TC intensity or structural metrics in DL models. Nonetheless, the low RMSE across metrics in both the H02H02 and L18L18 experiments highlights the TCNN model’s capability in retrieving TC intensity and structure on its own grid. This provides a foundation for further investigation into the effects of fine-scale processes in the next section.

\subsection{Fine-scale effects}
%
%
\begin{figure}[ht]
\centering
\includegraphics[width=13.5cm]{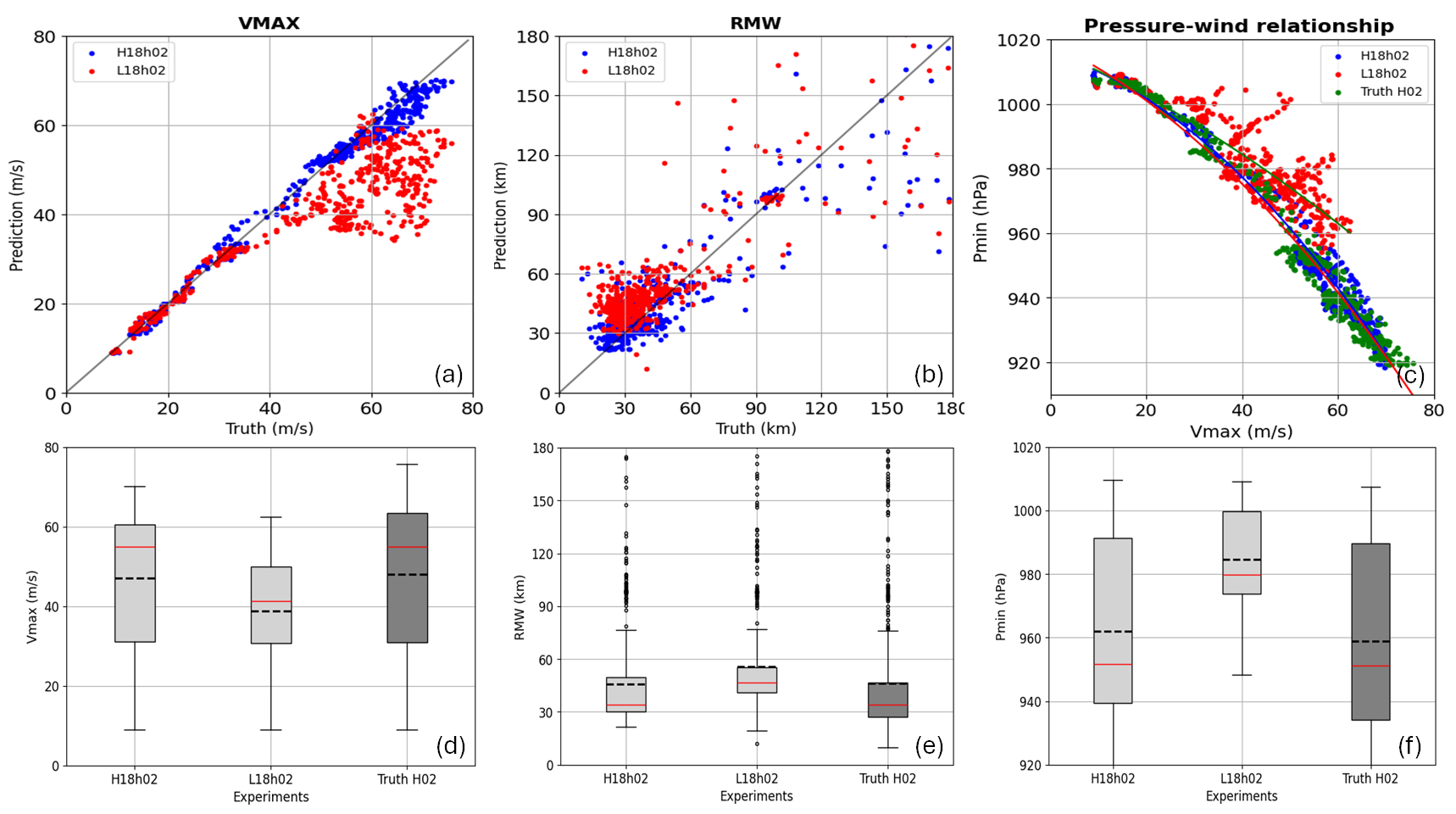}
\caption{a) A scatter plot of VMAX prediction for the H18h02 (blue) and L18h02 (red) experiments as obtained from a test set, b) Similar to (a) but for RMW prediction, c) the corresponding pressure-wind relationships obtained from the H18h02 and L18h02 experiment, along with the true pressure-wind relationship obtained directly from the H02 simulations (green). (d)-(f) Box plots of the TCNN model prediction obtained from the test set, along with the true distribution, in the H18h02 and L18h02 experiments for VMAX (m s$^{-1}$, RMW (km), and PMIN (hPa), respectively.}
\label{fig:resolution-effect}
\end{figure}

Given that the TCNN model effectively retrieves TC intensity and structure on its own grid, the next question is whether it can capture TC information beyond its own grid. This is a crucial consideration for applying DL models to TC forecasting and downscaling, as all currently observed TC data exist independently of and outside the model grid of any current climate dataset. 

Our simulations in this study are designed to address this question in the clearest possible way. Specifically, we have two distinct 18-km datasets: the L18 dataset, which excludes feedback from fine-scale processes, and the H02 dataset, which incorporates such feedback. As emphasized earlier, the imprint of fine-scale processes on the coarser-resolution grid in the H02 dataset differs significantly from typical coarse/low-resolution images. This distinction arises because the two-way feedback in WRF simulations allows for multi-scale interactions that influence TC development, leading to variations in intensity and structure beyond the mere details observed at higher resolutions. By training and evaluating the TCNN model on both the L18 and H02 datasets, we can assess the capability of DL models in predicting extreme events at small spatial and temporal scales for practical applications.

For this, we conduct two DL experiments using the 18-km data from the L18 and H02 dataset. In one experiment, the 18-km data from the L18 dataset is used as the training dataset, while the target values are obtained from the 2-km resolution data from the H02 dataset. This experiment (L18h02) answers a question of how much unknown TC information at a finer resolution the TCNN model can retrieve from coarse-resolution simulations, given the same large-scale environmental setting. 

In the second experiment (H18h02), we use the 18-km data from H02 simulations for training but 2-km resolution data for target intensity and structure values. By comparing this H18h02 with the L18h02 experiment, one can examine how the manifestation of fine-scale processes on the coarser-grid affects DL models, which fundamentally differ from extrapolating detailed features from a coarse-resolution image in traditional image processing problems.  

Fig. \ref{fig:resolution-effect} shows the TCNN model's performance for these two DL experiments H18h02 and L18h02. One notices clearly the effect of fine-scale processes from these results across intensity and structure metrics. Without fine-scale effects, the TCNN model has RMSE errors of 12.72 m s$^{-1}$ for VMAX, 32.21 hPa for PMIN, and 20.15 km for RMW in the L18h02. The PWR is also significantly worse, with much of the PMIN higher than the true values at a given VMAX. 

Including fine-scale effects on the 18-km grid in the H18h02 experiment shows drastic improvement, with the RMSE for VMAX, PMIN, and RMW of 4.48 m s$^{-1}$, 5.56 hPa, and 17.22 km, respectively. A big improvement in all statistical measures and dynamical constraints is also observed in the box plots and scatter plots in Fig. \ref{fig:resolution-effect}. These improvements are non-trivial, as we recall that the TCNN model use 18-km grid data to retrieve TC information on an unseen higher resolution 2-km grid for both the H18h02 and L18h02 experiments. 

Apparently, the finer-scale effects are crucial for retrieving TC information correctly, as seen from these results. While L18h18 could perform well on its own 18-km data grid, using the unknown targets from 2-km resolution downgrades the model performance substantially. Such a downgraded performance is partly due not only to the resolution but also to the lack of the feedback from 2-km to 18-km resolution data as confirmed in the H18h02 experiment. 

The implication of this result is significant. First, it immediately indicates that the use of, e.g., 0.25-degree ERA5 data for training AI models, would be insufficient to capture strong TC intensity or structure due to the lack of fine-scale processes, which could feedback to the coarse-resolution grid and leave some imprints that DL models can capture if properly included. 

Second, these results suggest that the use of the NWP model to generate high-resolution data is an effective approach that can help capture high-resolution features that other approaches such as generative adversarial or diffusion could not obtain alone. This is because physical-based models can include some multi-scale interactions beyond just image enhancement. 

With the help of high-resolution dynamical downscaling using physical-based models to generate more data, one can therefore improve DL models for TC intensity and structure retrieval by taking into account processes that are otherwise absent in the coarse resolution climate reanalysis data. In this regard, our results indicate that a proper DL model and enhanced data from high-resolution NWP models can potentially help predict or downscale extreme weather, even when these extreme values are not on the model grid points. 

\subsection{Feature importance}
\begin{figure}[ht]
\centering
\includegraphics[width=12cm]{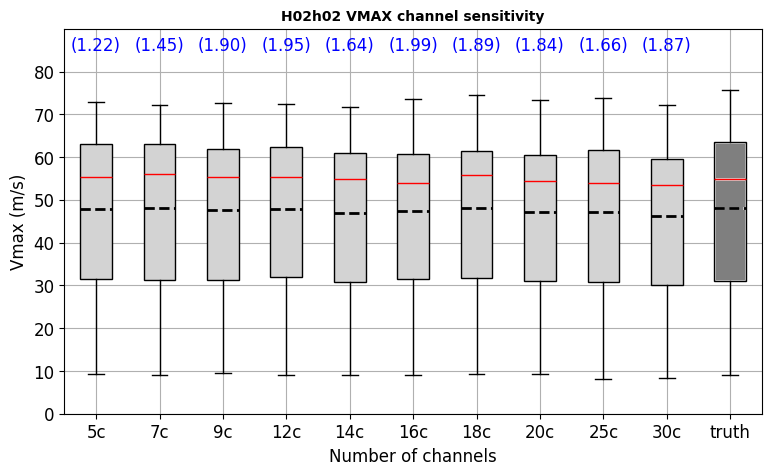}
\caption{Box plot distributions of VMAX prediction as obtained from the TCNN model for a test set in the H02h02 experiment, using increasing number of input channels. Blue numbers denote the root mean squared errors of VMAX prediction relative to the true distribution. The dark column denotes the true distribution of VMAX as obtained directly from the 2-km output of H02 simulations.}
\label{fig:channel_sensitivity}
\end{figure}

\begin{figure}[ht]
\centering
\includegraphics[width=12.5cm]{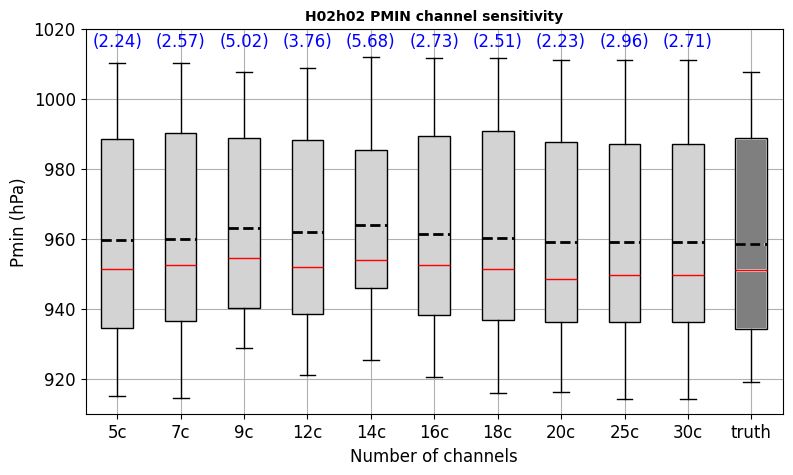}
\caption{Similar to Fig. \ref{fig:channel_sensitivity} but for PMIN.} 
\label{fig:Pmin_channel_sensitivity}
\end{figure}

\begin{figure}[ht]
\centering
\includegraphics[width=12.5cm]{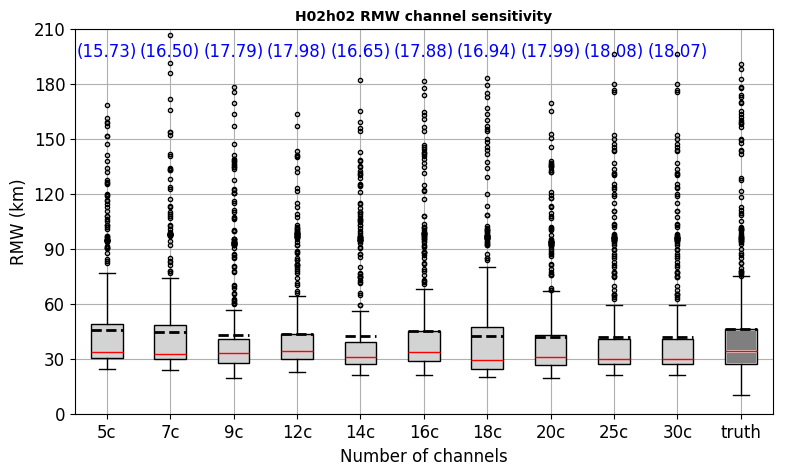}
\caption{Similar to Fig. \ref{fig:Pmin_channel_sensitivity} but for the RMW (km).}
\label{fig:Rmw_channel_sensitivity}
\end{figure}

The strong performance of the TCNN model in retrieving TC intensity and structure as demonstrated in the H02H02 and L18L18 experiments presents an opportunity for us to further investigate the key factors contributing to its success. Selecting input features for a DL model is often a trial-and-error process. For TC intensity and structure retrieval, there is no clear consensus on which meteorological variables should be minimally included for effective downscaling or forecasting. Previous theoretical models for TC development showed that TCs have unique structure and properties, with strong constraints among their dynamical and thermodynamic variables \cite{HackSchubert1986, Emanuel1988, Kieu2015, KieuWang2017a}. These constraints explain the well-defined PWR shown in Fig. \ref{fig:pminRmw} and suggest that one does not need to use too many channels to retrieve TC intensity and structure. This underlines our rationale for using just five channels in the TCNN model, which indeed suffices to retrieve TC intensity and structure as demonstrated in the previous section.     
In this section, we examine this channel data impacts by leveraging data from our 2-km WRF simulations and adding more input data to see if more channels can help improve TC retrieval. Figure \ref{fig:channel_sensitivity} shows the distributions and RMSE of VMAX prediction from the TCNN model using a range of new input channels shown in Table \ref{tab:channel}. A clear pattern emerges from this result: increasing the number of input channels from 5 to 30 generally degrades the model performance in some aspects such as RMSE, mean, or interquartile range as compared to the true distribution. For other statistical measures such as upper quartile, median or other intensity metrics such as PMIN or RMW (see Figs.\ref{fig:Pmin_channel_sensitivity}-\ref{fig:Rmw_channel_sensitivity} ), the model does not capture any significant changes between 5-18 channels. For more than 20 channels, RMSE and all statistical measures/metrics begin to increase or drift away from the true distributions, indicating the downgraded performance of the TCNN model when too many channels are included. 
%
%
\begin{table}[ht]
\centering
\caption{A list of channels used for training the TCNN model and its sensitivity analyses. By convention, a channel is denoted as $A_{z}$ where $A \in \{U, V, W, T, P, H, Q_v\}$ is one of 7 diagnostic variables in the WRF model, and $z$ is the height level in km. }\label{tab:channel}
\begin{tabular}{| p{1.3cm} | p{7.5cm} | p{3.5cm} |}
\hline
Acronym & Variables/Levels (km) & Remarks \\
\hline
05c & $U_1, V_1, T_8, Q_{v3}, P_0$ & Control settings with minimal input channels including near surface wind/moisture and upper-level temperature.\\
07c & $U_1, U_2, V_1, V_2, T_8, Q_{v3}, P_1$ & Add more near surface information. \\
09c & $U_1, U_2, U_{10}, V_1, V_2, V_{10}, T_1, T_8, T_9$ & Surface information with upper level wind components but no moisture, vertical motion, and pressure information   \\
12c & $U_1, U_2, U_{10}, V_1, V_2, V_{10}, W_8, W_9,$ $T_1, T_8, T_9, P_1$ & Surface information with upper level wind components but no moisture information.  \\
14c & $U_1, U_2, U_{10}, V_1, V_2, V_{10}, T_1, T_8, T_9, W_9$ $Q_{v1}, Q_{v2}, Q_{v3}, P_1$ &  All surface information, upper level wind components and moisture included but no vertical motion information.  \\
16c & $U_1, U_2, U_{10}, V_1, V_2, V_{10}, W_9, T_1, T_8, T_9,$ $Q_{v1}, Q_{v2}, Q_{v3}, H_1, H_5, P_1$ &  All surface information, upper level wind components, and moisture included. \\
18c & $U_1, U_2, U_{10}, U_{11}, V_1, V_2, V_{10}, V_{11}, W_9,$ $T_1, T_8, T_9, Q_{v1}, Q_{v2}, Q_{v3}, H_1, H_5, P_1$ & Similar to the control setting 16c but including 2 more upper level wind components. \\
20c & $U_1, U_2, U_{10}, U_{11}, V_1, V_2, V_{10}, V_{11}, W_9, H_1, H_5,$ $T_1, T_2, T_3, T_8, T_9, Q_{v1}, Q_{v2}, Q_{v3}, P_1$ & Similar to 18c but including more low-level temperature. \\
25c & $U_1, U_2, U_3, U_5, U_{10}, U_{11}, V_1, V_2, V_3, V_5, V_{10}, H_1, H_5$ $V_{11}, W_9, T_1, T_2, T_3, T_8, T_9, Q_{v1}, Q_{v2}, Q_{v3}, Q_{v8}, P_1$ & Similar to 20c but including more midlevel wind and upper-level moisture information.  \\
30c & $U_1, U_2, U_3, U_5, U_{10}, U_{11}, V_1, V_2, V_3, V_5, V_{10},H_1, H_5$ $V_{11}, W_1, W_5, W_6, W_7, W_9, T_1, T_2, T_3, T_8, T_9,$ $Q_{v1}, Q_{v2}, Q_{v3}, Q_{v8}, P_1$ & Similar to 25c but including more vertical wind information.  \\
\hline
\end{tabular}
\end{table} 

Among various channels, we notice that key contributors to the TCNN model performance are low-level horizontal winds/moisture and mid-to-upper-level temperature. Adding more channels beyond these essential channels does not help improve the model further. In fact, excessive input channels degrade the model performance by introducing redundant information, leading to model confusion rather than improvement as noted in \cite{NguyenKieu2024}. With the high-resolution data generated from the WRF model, we thus not only identify the most effective channels for retrieving TC information but also confirm the strong constraints on TC dynamics previously established by theoretical and numerical studies, using our data-driven TCNN model.
%
%
%
%
%
%
%
%
\section{Discussion and Conclusion}\label{sec:conclusion}
In this study, we proposed a new approach that can improve the application of DL to TC intensity downscaling from gridded climate or weather data. Using a numerical weather prediction (NWP) model with different model physics to generate high-resolution data for TC structure, we showed that fine-scaled processes have significant effects on the larger-scale environment for which DL models can capture. Unlike the typical fine-grain and coarse-grain features in image processing, our results showed that the multi-scale interaction between TC and environment can leave some imprints on the storm-scale dynamics and structures. With a properly designed DL model, these imprints can be captured and help enhance the accuracy of TC intensity downscaling across metrics.

The implications of these results are important in several aspects. First, we demonstrated that the lack of accounting for fine-scale processes in current climate reanalysis datasets such as ERA5 is the main reason why global AI models could perform well for large-scale metrics or storm track, but cannot compare to regional physical-based models in predicting TC intensity. Regardless of the algorithms that can help enhance TC structure and intensity from these coarse-resolution climate data, not accounting for the TC-environmental interaction between convective scale and storm-scale would limit the performance of global AI models in predicting TC intensity in the future.

Second, our results present a broader approach for applying DL methods for extreme weather beyond TC intensity. Specifically, we can use NWP models as a data generator for each type of extreme weather event that helps train a DL model in a similar way as a generative adversarial network. The need for a good NWP model is crucial here, as it plays the role of produce "unseen" high-resolution structures beyond what our current observation network can afford. Of course, one can always attempt to develop a new observing system that captures three-dimensional structures and the intensity of any weather event. However, carrying out and maintaining such observing networks for a set of extreme weather events is much more impractical than using NWP models to do the same job. Although NWP models are not perfect, especially when simulating extreme events, they are the most practical and increasingly useful when more knowledge and understanding of extreme events are further incorporated. From this perspective, we support of maintaining and continuous developing NWP models as an inseparable component for AI/ML applications in extreme weather prediction.   

While the results presented in this study could highlight several important issues and directions for applying AI models at the regional level, a few shortcomings should also be mentioned. First, our DL model and results are specifically tuned for the WRF model output at a resolution of 2 km. How much our results would change for even higher resolutions or other modeling systems are still elusive. Answering this question require much more intensive experiments and training beyond our current capability. Note that this drawback will not change the value or significance of our approach herein, but it could impose some limits on the performance of our DL model or the robustness of the results in this study.

Second, the idealized setting in our WRF simulation is good for quantifying a clean impact of model resolution, yet it does not allow for more realistic environmental control or land-sea effects. In practice, TCs constantly move from one place to the other and so any regional AI models should track TC motion and predict TC intensity within a storm-following framework. This problem is a difficult one in practice, as tracking a TC in a regional model would require a guiding model, which may impose the wrong environment for TC motion. Moreover, training a DL model to downscale TC intensity for these moving systems becomes harder, as we have to take into account lateral boundary conditions that our design in this study could not capture.     

\section*{Acknowledgments}
This research was funded by the NSF (AGS \# 2309929).

\section*{Author contribution} 
CK perceived the ideas, designed the workflow, analyzed the results, and wrote the draft of this work. KL and TN built models, conducted experiments, and helped with data visualization and analyses.

\bibliographystyle{unsrtnat}
\bibliography{references1,references2}  






\end{document}